\documentclass[conference]{IEEEtran}
\IEEEoverridecommandlockouts
\usepackage{cite}
\usepackage{amsmath,amssymb,amsfonts}
\usepackage{algorithmic}
\usepackage{graphicx}
\usepackage{textcomp}
\usepackage{xcolor}
\usepackage{graphicx}
\usepackage{float} 
\usepackage{caption}
\usepackage{subfigure}
\usepackage{amsfonts}
\usepackage{amsmath}
\usepackage{graphicx} 
\usepackage[colorlinks]{hyperref}
\usepackage{microtype}
\usepackage[normalem]{ulem}
\usepackage{array}
\usepackage{multirow}
\usepackage{amsmath}
\usepackage{booktabs}
\usepackage{tabularx}
\usepackage{stfloats}
\usepackage[top=0.7in, left=0.8in, bottom=0.95in, right=0.8in]{geometry}
\renewcommand*{\thesubsection}{\alph{subsection}}
\def\BibTeX{{\rm B\kern-.05em{\sc i\kern-.025em b}\kern-.08em
    T\kern-.1667em\lower.7ex\hbox{E}\kern-.125emX}}

\begin{document}

\title{Efficient Autoprecoder-based deep learning for massive MU-MIMO Downlink under PA Non-Linearities
}

\author{\IEEEauthorblockN{Xinying Cheng$^1$,  Rafik Zayani$^2$, Marin Ferecatu$^1$, Nicolas Audebert$^1$ \\ $^1$CEDRIC (EA4329), Conservatoire national des arts et métiers, Paris 75003, France \\
$^2$CEA-Leti, Univ. Grenoble Alpes, F-38000 Grenoble, France\\
}}

\maketitle

\begin{abstract}
This paper introduces a new efficient autoprecoder (AP) based deep learning approach  for massive multiple-input multiple-output (mMIMO) downlink systems in which the base station is equipped with a large number of antennas with energy-efficient power amplifiers (PAs) and serves multiple user terminals. We present AP-mMIMO, a new method that jointly eliminates the multi-user interference and compensates the severe nonlinear (NL) PA distortions. Unlike previous works, AP-mMIMO has a low computational complexity, making it suitable for a global energy-efficient system.
Specifically, we aim to design the PA-aware precoder and the receive decoder by leveraging the concept of \emph{autoprecoder}, whereas the end-to-end massive multi-user (MU)-MIMO downlink is designed using a deep neural network (NN). 
Most importantly, the proposed AP-mMIMO is suited for the varying block fading channel scenario. To deal with such scenarios, we consider a two-stage precoding scheme: 1) a NN-precoder is used to address the PA non-linearities and 2) a linear precoder is used to suppress the multi-user interference. The NN-precoder and the receive decoder are trained off-line and when the channel varies, only the linear precoder changes on-line. This latter is designed by using the widely used zero-forcing precoding scheme or its low-complexity version based on matrix polynomials. 
Numerical simulations show that the proposed AP-mMIMO approach achieves competitive performance with a significantly lower complexity compared to existing literature.
\end{abstract}


\section{Introduction}
\label{sec:intro}

The future wireless networks will be able to accommodate a broad collection of facilities of varying specifications, with massive multiple-input multiple-output (mMIMO) technology emerging as a core feature. Usually, direct linear signal processing methods, such as the matched filter and zero-forcing (ZF), are widely used as precoding techniques for mMIMO downlinks, because of their ability to reduce transmitter power consumption, to realize signal transmission and to eliminate multi-user interference \cite{wang2017zf}. These linear precoders, on the other hand, carry a high peak-to-average power ratio (PAPR) to transmit signals \cite{guerreiro2016massive}. As a consequence of the high PAPR, mMIMO signals are very sensitive to the non-linearity of the radio frequency power amplifier (PA), which is the main hardware distortion and is inevitable in a typical transmission chain. Indeed, when PA is present in the transmission chain, signal distortion and phase rotation emerge, resulting in a transmission performance degradation.

There are a variety of methods that can be used to reduce the nonlinear distortion described above. One solution is to apply a back-off to the power amplifier input signal preventing nonlinear-operation of the PA; however, this implementation reduces the PA efficiency. Another classical solution is the digital predistortion (DPD), which is inserted before the power amplifier stage to compensate its nonlinear behaviors \cite{kim2005digital}. Although DPD allows more effective linear operation of the PAs, classic DPD techniques may be deemed unsuitable for mMIMO due to the significant additional complexity. Brihuega \emph{et al.} \cite{brihuega2018digital} explored a DPD method based on minimizing the nonlinear distortion of the combined signal at the intended receiver direction, and proposed a low-complexity version to linearize an arbitrary number of parallel and mutually different power amplifiers, with only a single DPD. However, this method still has a high computational complexity and it is limited to the context of single-user fully digital beamforming transmitters, which is not suitable for mMIMO scenario. 

There are other methods that aim to decrease the PAPR of the transmitted signal for the mMIMO scenario, the motivation being that a low PAPR can avoid the PA nonlinearity when PA is operated with high efficiency. In this sense, \cite{siegl2011selected} and \cite{chen2017low} proposed several optimisation algorithms focusing on reducing the PAPR in MU-MIMO systems. Instead of solving the convex problem, another algorithm-based solution is proposed in \cite{yao2018semidefinite}, in which the author uses relaxation of the nonconvex constraints to obtain a convex problem that approximates the original problem. In the same direction, Zayani \emph{et al.} have proposed the MU-PNL-GDm algorithm based on gradient descent, performing jointly linear precoding, PAPR reduction and DPD but it also suffers of great computational complexity and high latency since it consists on an iterative algorithm based gradient descent approach \cite{zayani2019efficient}.

On the other hand, machine learning methods, and recently deep learning (DL), have been successfully applied to efficiently perform the real-time optimization tasks required for massive MU-MIMO. Sohrabi et al. \cite{sohrabi2019one} proposed an autoencoder to enhance the performance of one-bit massive MIMO systems: this approach has been shown to achieve satisfactory performance but it is limited to multicasting communications systems. Moreover, its computational complexity is still challenging due to the high size of the deployed neural networks. It is worth mentioning that the concept of end-to-end learning based autoencoder has been widely studied for single-input single-output (SISO) systems and has been shown to deal with  hardware imperfections \cite{dorner2017deep} \cite{felix2018ofdm}. To the best of our knowledge, none of existing works has explored the potential of DL in dealing with PA non-linearities of massive MU-MIMO Downlink systems over varying fading channels.

Regarding the issues discussed above and the motivations behind this work, our main contributions are summarized as follows:
\begin{itemize}
    \item  We present the first successful exploration of the potential of DL in dealing with PA non-linearities in massive MU-MIMO Downlink systems over varying fading channels. The key idea is to represent transmit precoder, PAs, channel and receiver as one deep NN, i.e., as an end-to-end reconstruction task where only the transmit precoder and the receive decoder are trainable.      
    
    \item  Unlike the existing literature \cite{sohrabi2019one}\cite{dorner2017deep}\cite{felix2018ofdm}, our proposed AP-mMIMO is suited for varying channel scenario, i.e., its adaptation requires lower computational complexity. For this, we consider a two-stage precoding scheme : (1) a \emph{NN-precoder}, which is trained off-line and fits for any channel, and (2) a \emph{Linear precoder} that depends on the channel and can be designed using the widely used zero-forcing precoding scheme or its low-complexity MP-based version. 
    
    \item In order to avoid the high computational complexity entailed by the large-scale matrix inversion required for ZF precoding, we adopt matrix polynomials (MP) for data precoding \cite{zhu2015linear} in the second stage precoding.  Numerical simulations show that these latter can achieve performance close to that of the ZF precoder. It is worth to mention that, unlike \cite{sohrabi2019one}, the user power allocation in our proposed AP-mMIMO is straightforward.
\end{itemize}

The rest of this paper is structured as follows. Section~\ref{sec:system} presents the system model investigated in this study. Section~\ref{sec:AP} introduces the proposed AP-mMIMO framework and the learning procedure. In Section~\ref{sec:complexity}, we investigate the complexity of the proposed solution and present comparisons to other approaches from the literature. Finally, numerical simulation results are provided in Section~\ref{sec:results} and the paper concludes in Section~\ref{sec:conclu} with a discussion of open challenges and areas for future investigations.

\emph{Notations}: Lower-case letters (e.g. $x$) , bold lower-case letters (e.g. $\mathbf{x}$) and bold upper-case letters (e.g. $\mathbf{X}$) stand for scalars, vectors and matrices, respectively. We denote the transpose and conjugate transpose by $\mathbf{X}^T$ and $\mathbf{X}^H$, respectively.  We use $\parallel\mathbf{x}\parallel_2$ and $\mathbb{E} \{ . \}$ to denote $l_2$-norm of vector $\mathbf{x}$ and the expectation operation, respectively.

\begin{figure*}
\centerline{\includegraphics[width=\textwidth]{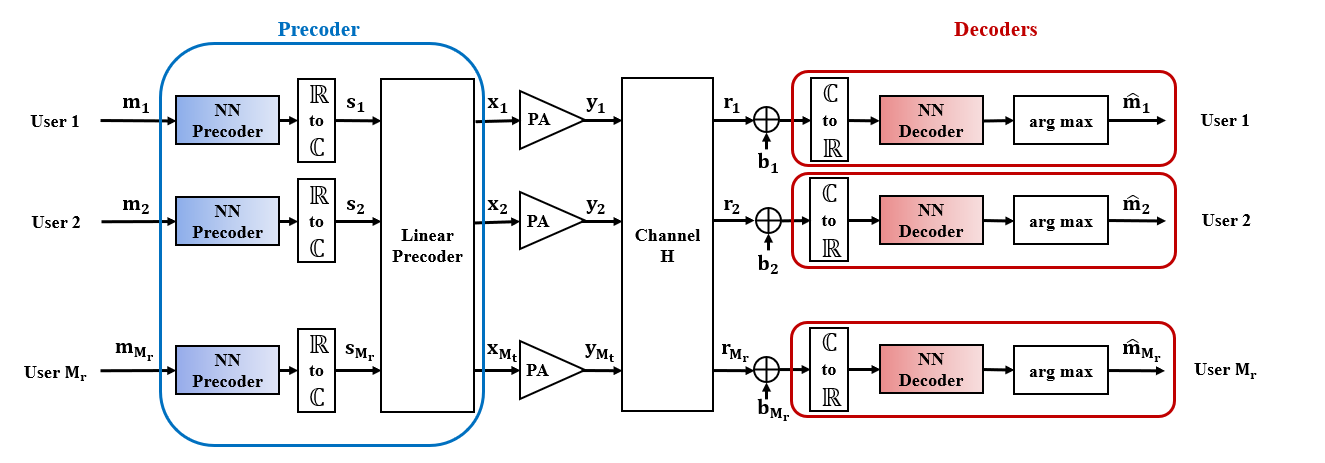}}
\caption{Proposed AP-mMIMO structure: Generalization phase.}
\label{fig2}
\end{figure*}

\section{Massive MU-MIMO signal and Nonlinear PA impairment model}
\label{sec:system}

We consider, in this paper, a single-cell massive MIMO-OFDM downlink system, where $M_r$ single-antenna users are served simultaneously by the base station (BS) equipped with $M_t$ antennas, over a fading channel $\mathbf{H} \in \mathbb{C}^{M_r\times M_t}$ which has random complex Gaussian entries. $M_t$ is significantly larger than $M_r$. The symbols for the $M_r$ users are included in the $\mathbb{C}^{M_r\times1} $ signal vector $\mathbf{s}$, where $s_{m_r}$ is chosen from a $M$-quadrature amplitude modulation (QAM) constellation, where $M$ is the modulation order. The BS uses a linear precoding stage, transforming the vector $\mathbf{s}$ into a $M_t$-dimensional vector by using a $\mathbb{C}^{M_t\times M_r} $ matrix $\mathbf{W}=\left[\mathbf{w}_{1}, \mathbf{w}_{2}, \ldots, \mathbf{w}_{M_{r}}\right]$, using the channel state information (CSI). This can be written as

\begin{equation}
\mathbf{x}=\frac{1}{\sqrt{\varsigma_W}} \mathbf{W s} \label{eq:ZF}
\end{equation}
where $\frac{1}{\sqrt{\varsigma_W}}$ is the power normalization factor.
Here, the precoding matrix can be obtained by using the ZF technique \cite{spencer2004zero} or by the matrix polynomials technique \cite{zhu2015linear}, which are calculated by $\mathbf{W}=\mathbf{H}^{H}\left(\mathbf{H} \mathbf{H}^{H}\right)^{-1}$ and $\mathbf{W} =  \mathbf{H}^{H} \sum_{j=0}^{\mathcal{J}} \mu_{j}\left( \mathbf{H} \mathbf{H}^{H}\right)^{j}$, respectively, where $\mathbf{\mu}$ is a vector containing the real-valued coefficients of the precoder matrix polynomial. The factors $\mathbf{\mu}$ can be optimized theoretically and they do not depend on the instantaneous channel estimates. Note that the Horner's implementation rule \cite{zhu2015linear} should be adopted for seeking low complexity. Interested readers are referred to \cite{zhu2015linear} for more details concerning the $\mathbf{\mu}$ coefficients optimization and the Horner's rule implementation.    

For the case of massive MIMO with non-linear PAs, the precoded symbols in Equation (\ref{eq:ZF}) are fed, towards the BS antennas, through $M_t$ parallel transmit chains with PAs (see Fig.~\ref{fig2}). The resulting amplified symbols are
\begin{equation}
\mathbf{y}=\left[f_{1}\left(x_{1}\right), f_{2}\left(x_{2}\right), \ldots, f_{M_{t}}\left(x_{M_{t}}\right)\right]^{T}=F(\mathbf{x})
\end{equation}
where $f_{m_t}(.)$ denotes the nonlinear amplification operation of the $m_t$-th PA. Then, the symbols received by the $M_r$ users can be written as 
\begin{equation}
\mathbf{r} = \mathbf{H} \mathbf{y} + \mathbf{b} \label{eq:transmission}
\end{equation}
where $ \textbf {b} $ is the added Gaussian noise vector, and its entries are i.i.d random variables having a circularly-symmetric distribution with zero mean and $ \sigma_b^2 $ variance.

For our MIMO system, the major sources of distortions in the transmitter are the nonlinear PAs, especially when they are operated close to their saturation regions to increase their power-efficiency. Now, we let $\mathbf{x}_{m_t}(n) = \rho(n) e^{j\phi(n)}$ be the $n$-th sampling point that is to be transmitted and amplified via the antenna $m_t$, where $\rho(n)$ and $\phi(n)$ denote, respectively, the magnitude and phase of that sample. Then, the relation between the baseband equivalent input and the output signals of the PA in the $m_t$-th antenna branch can be written as

\begin{equation}
\mathbf{y}_{m_t}(n) = g(\alpha \rho(n)) e^{j(\phi+\Psi(\alpha \rho(n)))}
\end{equation}
where $g(.)$ is the amplitude-to-amplitude (AM-AM) conversion and $\Psi(.)$ is the amplitude-to-phase (AM-PM) conversion of the PA. The factor $\alpha$ is a multiplicative coefficient applied at the input of the PA to obtain the PA working point according to the given input back-off (IBO). Here, all IBOs are given in dB related to its saturation point. The factor $\alpha$ is needed to ensure a signal $\mathbf{x}_{m_t}(n)$ with a given IBO value, and is obtained through $\sqrt{{P_{sat}} / (10^{\frac{I B O}{10}} P_{t}})$. $P_{sat}$ is the saturation input power of the PA and $P_{t}$ represents the transmitting power of the signal. In this paper, the functions $g(\rho)$ and $\Psi(\rho)$ are modeled by the modified Rapp model proposed by the 3GPP for the New Radio (NR) evaluation \cite{rapp}. In this modified Rapp, which reflects closely to realistic PAs, AM/AM and AM/PM conversions can be described by

\begin{equation}
g(\rho)=\frac{G \rho}{\left(1+\left|\frac{G \rho}{V_{s a t}}\right|^{2 p}\right)^{\frac{1}{2 p}}}, \quad \Psi(\rho)=\frac{A \rho^{q}}{\left(1+\left(\frac{\rho}{B}\right)^{q}\right)}
\end{equation}
where $G$ is the small signal gain, $V_{sat}$ is the saturation level, $p$ is the smoothness factor and $A$, $B$ and $q$ are fitting parameters.

\section{Proposed Autoprecoder structure and learning solution}
\label{sec:AP}

In this section, we present our autoprecoder based end-to-end mMIMO communication structure and the training/generalization details.
\subsection{Autoprecoder structure representation}

The structure of the proposed AP-mMIMO is shown in Fig.~\ref{fig1}.
Since most deep learning libraries currently support real-valued operations, we map complex vectors to real ones by concatenation of real and imaginary parts ($\mathbb{R}$-to-$\mathbb{C}$ and $\mathbb{C}$-to-$\mathbb{R}$ blocks in Fig.~\ref{fig1}).


The symbol ${c} \in \{1,...,M\}$ denotes the index of the intended message for a given user and $\mathbf{m}$ is its one-hot representation, i.e., a $M$-dimensional vector with the $m$-th element being one and the other elements being zero. This vector is fed to the transmit neural network with multiple dense nonlinear layers followed by a linear layer. Here, a dense layer is a matrix multiplication with learnable weights followed by a non-linear activation function. The linear layer should contain two linear neurons representing the outputs $I$ and $Q$ of the base-band complex signal corresponding to one user.

The $2$-dimensional real-valued signal at the output of the NN-precoder $\widetilde{\mathbf{m}}$ in this model can be written as
\begin{equation}
\widetilde{\mathbf{m}}=\mathbf{K}_{\mathrm{T}}  \cdot \sigma_{\mathrm{T}-1}\left(\cdots \mathbf{K}_{2}  \cdot \sigma_{1}\left(\mathbf{K}_{1}   \cdot \mathbf{1}_{m}+\mathbf{n}_{1}\right)+\cdots \mathbf{n}_{\mathrm{T}-1}\right)+\mathbf{n}_{\mathrm{T}}
\end{equation}
where $\mathrm{T}$ is the number of layers (i.e., $\mathrm{T-1}$ hidden layers and one output layer), $\sigma_t$ is the activation function for the $t$-th layer. $\mathbf{\Theta}_{\mathrm{T}}=\left\{\mathbf{K}_{\mathrm{t}}, \mathbf{n}_{\mathrm{t}}\right\}_{\mathrm{t}=1}^{\mathrm{T}}$ represents the set of the NN parameters (matrices and bias). The dimensions of the NN parameters are 
\begin{equation}
\begin{array}{l}
\operatorname{dim}\left(\mathbf{K}_{\mathrm{t}}\right)=\left\{\begin{array}{ll}
\ell_{\mathrm{t}} \times M, & t=1 \\
\ell_{\mathrm{t}} \times \ell_{\mathrm{t}-1}, & t=2, \ldots, T-1, \\
2 \times \ell_{\mathrm{t}-1}, & t=T,
\end{array}\right. \\
\operatorname{dim}\left(\mathbf{n}_{\mathrm{t}}\right)=\left\{\begin{array}{ll}
\ell_{\mathrm{t}} \times 1, & t=1, \ldots, T-1, \\
2 \times 1, & t=T
\end{array}\right.
\end{array}
\end{equation}
where $\ell_{\mathrm{t}}$ is the number of neurons in the $t$-th hidden layer.

Note that only one NN-precoder is trained during the learning phase (as shown by Fig.~\ref{fig1}) and is used for the $M_r$ users at test time. In the generalization phase, we generate the $2M_r$ real-valued signal corresponding to the $M_r$ users (as shown by Fig.~\ref{fig2}), performing as the first stage of precoding. The second stage is a classical linear precoder, e.g. zero-forcing one. It is worth noticing that the transmit signal, at each antenna, at the output of the linear precoder is adjusted to the desired IBO (for the sake of simplicity, it is not presented in the Figs.~\ref{fig2} and~\ref{fig1}).       

Similarly to the transmitting NN precoder, another NN with $R$ dense layers is used as decoder at the receiver side.  In the NN-decoder, the $r$-th layer contains $\ell_{\mathrm{r}}$ neurons,  models the receivers' operations. In our system the same NN-decoder is used for all user's decoding procedures. Similar as the precoder, only one NN-decoder is trained during the off-line learning phase, however all users used the same decoder during the on-line test phase. 

It is worth mentioning that by using this approach, we decrease the measurements of the receivers' training parameters, theoretically resulting in a faster training process and low-complexity precoder/decoder on-line running. Note that the last layer's activation function is set to softmax activation in the receivers' NN in order to produce the probability vector at user $m_r$, ${p}_{k} \in(0,1)^{|\mathcal{M}|}$, where the $i$-th element indicates the probability for the intended symbol's index to be $i$. Finally, user $m_r$ declares $\hat{m}$, as the decoded index of the intended symbol, which corresponds to the index of the element of ${p}_{k}$ with the highest probability. This is required so that we can train the network using cross-entropy to make the model learns to map the correct output with the correct input.

\subsection{Implementation details}

\begin{figure*}
\centerline{\includegraphics[width=\textwidth]{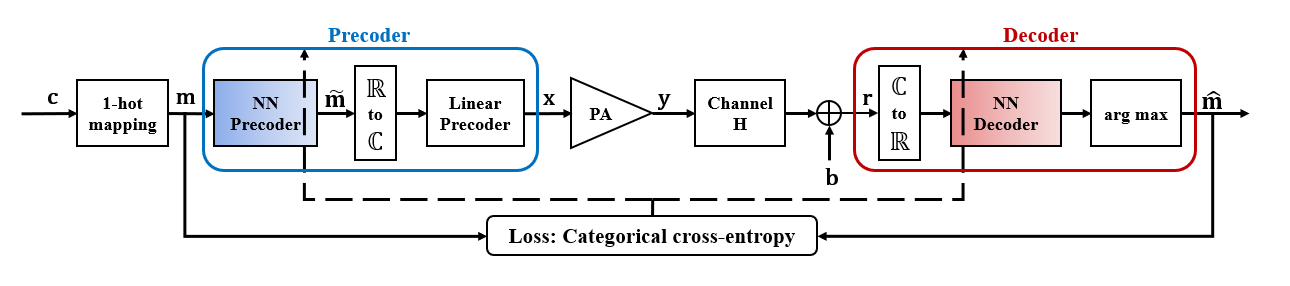}}
\caption{Structure of the learning phase. Two NNs are trained, one at the transmitter side and another one at the receiver side, using categorical cross-entropy as loss function.}
\label{fig1}
\end{figure*}

The AP-mMIMO based deep NN in Fig.~\ref{fig1} is implemented using TensorFlow, an open source Python-based machine learning framework. To train the NN-precoder and the NN-decoder, we use the Adamax optimizer, a version of the stochastic gradient descent approach for optimizing neural networks. We use the categorical cross-entropy as the loss function. Each of the NN-precoder and NN-decoder has only one hidden layer (i.e., $T=2$ and $R=2$) with, respectively, $\ell_{\mathrm{t}}$ and $\ell_{\mathrm{r}}$. 
Furthermore, we use rectified linear unit (ReLU) as the hidden layer activation function.

Fig.~\ref{fig2} illustrates the generalization phase of our proposed AP-mMIMO illustrating a typical massive MIMO downlink system. We recall that the NN-precoder and NN-decoder blocks are identical for all users and are trained with $10^5$ channel matrix $\mathbf{H}$, which are chosen randomly. For the proposed NN-precoder and NN-decoder, we generate training set $\textbf{m}$ with size of also $10^5$. 

Finally, we train both the NN-precoder and the NN-decoder, simultaneously and through the different mMIMO chain operations, using the categorical cross-entropy loss as shown in Fig.~\ref{fig1}. The cross-entropy constrains the output of the decoder to be the same symbol as the one in input to the precoder. Gradients are backpropagated through the decoder and then the precoder, which allows us to perform the weight updates \cite{DLBook2016}.

\section{Computational Complexity Analysis}
\label{sec:complexity}

\begin{table*}[t]
\newcommand{\tabincell}[2]{\begin{tabular}{@{}#1@{}}#2\end{tabular}}

\caption{Complexity analysis}

\begin{tabularx}{\textwidth}{X  c c c c }

\toprule
Method  & ZF+DPD & MU-PNL-GDm \cite{zayani2019efficient} & AP-mMIMO (ZF) & AP-mMIMO (MP)  \\
\midrule
Linear precoder computation & $8 M_t M_r^2 + M_r^3$ & - & $8 M_t M_r^2 + M_r^3$  & $-$ \\
\midrule
NN-precoder/decoder  &- &- & $(M+2)\ell_{\mathrm{t}} M_r \tau + (M+2)\ell_{\mathrm{r}} M_r \tau $ & $(M+2)\ell_{\mathrm{t}} M_r \tau + (M+2)\ell_{\mathrm{r}} M_r \tau $ \\
\midrule
Precoding vector update & $4 \tau M_t M_r$ & $\tau N_\text{iter} (12M_tM_r + 128M_t)$  & $4 \tau M_t M_r$ & $4 \tau (2 \mathcal{J}+1) M_t M_r$ \cite{zhu2015linear}  \\
\midrule
DPD & $\tau 128 M_t $ &- &- &- \\
\bottomrule
\end{tabularx}

\label{table:complexity}
\end{table*}

In this section, we analyze the online computational complexity of the proposed AP-mMIMO and compare it against the algorithm based solution MU-PNL-GDm \cite{zayani2019efficient}, which jointly perform linear precoding, PAPR reduction and DPD. To the best of our knowledge, among the existing literature, the MU-PNL-GDm offers the best performance in terms of symbol error rate (SER) but its computational complexity and latency are too high, making it not suitable for real-time massive MU-MIMO downlink systems.  

We recall that, in the proposed AP-mMIMO, the linear precoding can be either the ZF or its variant based on the matrix polynomials. Consequently, we have two AP-mMIMO schemes : 1) AP-mMIMO (ZF) and 2) AP-mMIMO (MP). 

Table ~\ref{table:complexity} summarizes the complexities of implementing the four different methods (i.e., AP-mMIMO (ZF), AP-mMIMO (MP), MU-PNL-GDm \cite{zayani2019efficient}) and classical ZF precoding and DPD combination method. In the last case, dedicated DPD block per antenna branch is used. This DPD is based on the well-known multi-layer perception (MLP) NN model. The NN predistorter has two inputs, two outputs and one hidden layer with $32$ neurons, demanding then $128$ real multiplication to execute one symbol. For all scenarios, we compute the number of real multiplications where $\tau$ represents the channel coherence interval, means that the channel characteristics keep constant during $\tau$ symbols. 

Note that the training of the NN-precoder and the NN-decoder is performed off-line, so its corresponding complexity is not considered. Therefore, the online computational complexity, during a channel coherence interval (i.e., processing $\tau$ symbols) , related to the proposed AP-mMIMO includes the performing of the NL precoding, linear precoding (ZF or MP) and NL decoding. For other methods, the total computational complexity includes linear precoder preparation and precoding vector update. Concerning the complexity of the MU-PNL-GDm, readers are referred to \cite{zayani2019efficient} for further details. Here, we note that the $N_{iter}$ refers to the number of total iterations in the algorithm in order to achieve certain performance. 

\section{Simulation Results}
\label{sec:results}

\begin{figure}
\centerline{\includegraphics[width=\columnwidth]{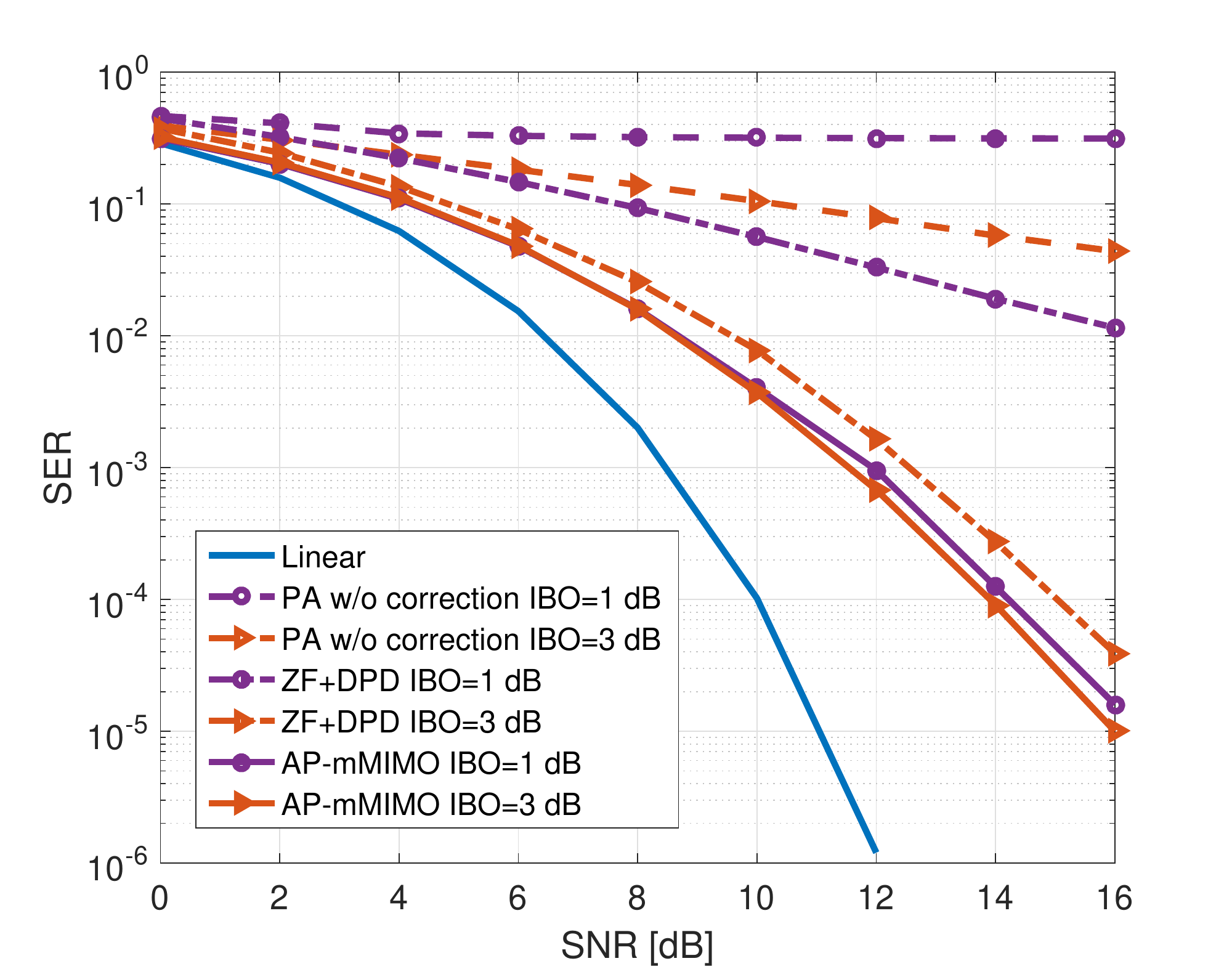}}
\caption{SER vs. SNR comparison when PA is operated at IBO = 1 dB and 3 dB }
\label{fig3}
\end{figure}

\begin{figure}
\centerline{\includegraphics[width=\columnwidth]{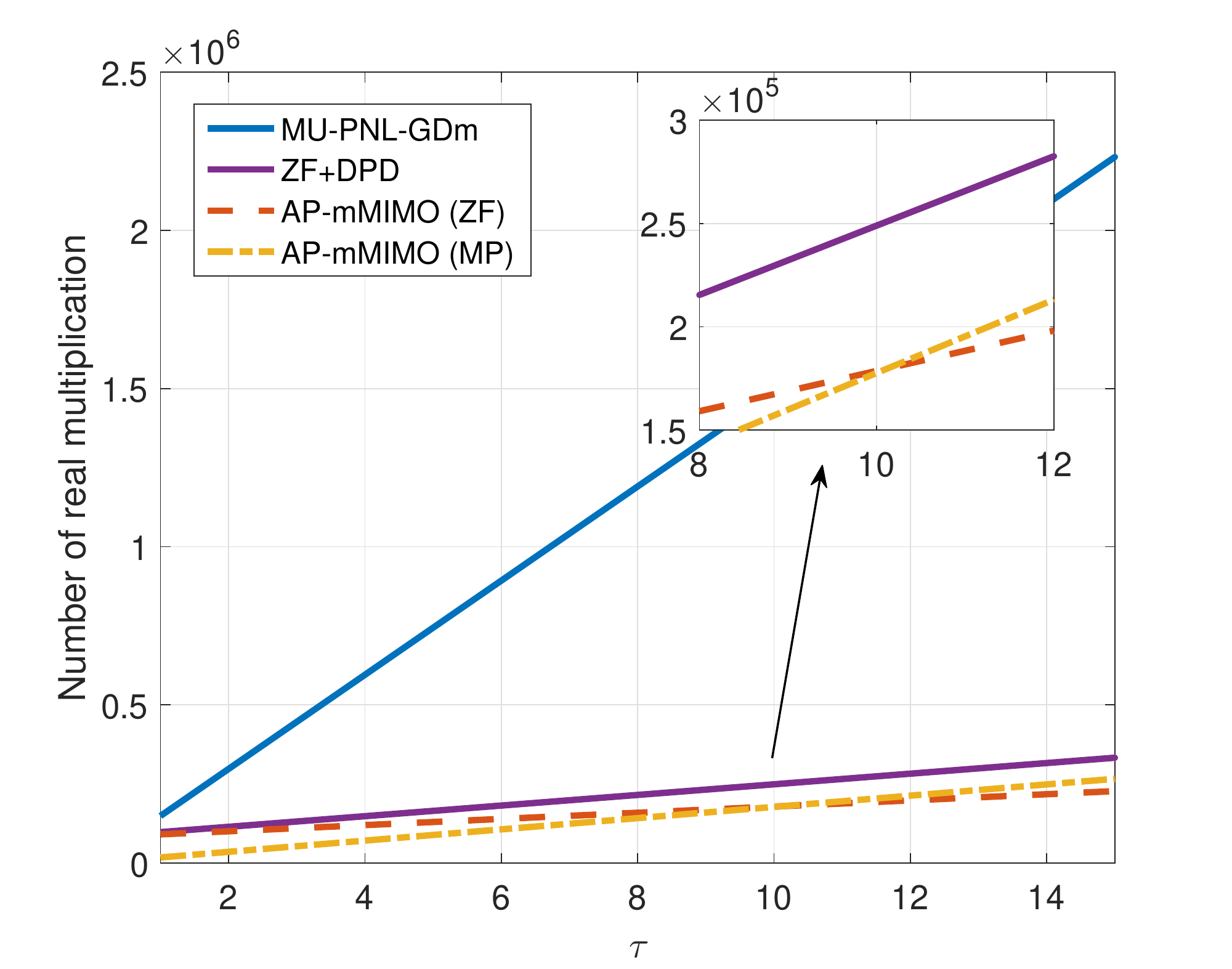}}
\caption{Complexity vs. $\tau$. }
\label{fig4}
\end{figure}

In this section, we evaluate the performance of the proposed AP-mMIMO based end-to-end massive MU-MIMO downlink systems under PA non-linearities. For all the simulations, we consider the BS to be equipped with $M_t=100$ antennas and serving $M_r = 10$ single-antenna users simultaneously. $16$-quadrature amplitude modulation (QAM) is considered (i.e., $M=16$). The memory-less modified Rapp model is adopted to determine the PA input and output, with parameters $G = 16$, $V_{sat} = 1.9$, $p = 1.1$, $A = -345$, $B = 0.17$, and $q = 4$ \cite{rapp}. Note that new test data over $10^5$ channel realizations were considered to evaluate the SER performance.

Fig.~\ref{fig3} illustrates the SER vs. signal-to-noise ratio (SNR) of the proposed AP-mMIMO when PAs are operating at IBO = $1$ dB and $3$ dB. As references, we plot the scenario that there is no PA distortion correction (``PA w/o correction'') and the linear case where the PA is considered as an ideal one (``Linear''). As comparison, we plot the scenario where ZF precoding and DPD method are applied in the communication system (``ZF+DPD''). We recall that, in this case, dedicated DPD block based on MLP NN model is used at each antenna branch. Finally, we plot the results of our proposed AP-mMIMO method in solid lines. Here, each of the NN-precoder and NN-decoder has one hidden layer of $16$ neurons and the SNR is defined as $\operatorname{SNR} = \mathbb{E} \{ \left\| y_{n} \right\|_{2}^{2} \} / {\sigma_b^2}$. From the Fig.~\ref{fig3}, we can clearly note that our proposed AP-mMIMO scheme provides significant improvements compared to the case without correction, which is quite close to the one provided by an ideal massive MU-MIMO system. With IBO=3 dB, the performance of AP-mMIMO is slightly better than that of the classical technique (ZF+DPD). However, this performance improvement becomes significant when IBO = 1 dB, i.e. when the PA is operating in a region with higher energy efficiency. It is worth mentioning that the two options of the proposed AP-mMIMO, either with ZF or MP, provide the same performance in terms of SER when $\mathcal{J}=5$, as mentioned in \cite{zhu2015linear}, where $\mathcal{J}=5$ represents the order of the polynomial. Then, for the sake of simplicity, only one AP-mMIMO SER curve is given for each IBO value.      

Most importantly, the advantage of our proposed method is the extreme low complexity needed in online mode, which makes it interesting for real-time massive MU-MIMO downlink systems. Indeed, according to the results illustrated in Fig.~\ref{fig4}, the computational complexity vs. $\tau$, one can note that our proposed method, the AP-mMIMO ZF/MP, has much lower complexity for all range of $\tau $ compared to the MU-PNL-GDm algorithm \cite{zayani2019efficient}. Compared with the classical ZF+DPD method, our proposed method has lower complexity while achieving better SER performance, especially when the power-efficiency is sufficiently high. This confirm the capability of such global optimization (end-to-end learning) in tackling PA non-linearity. Here, $N_{iter} = 6$ and $\mathcal{J}=5$, as presented, respectively, in \cite{zayani2019efficient} and \cite{zhu2015linear}. When $\tau = 5$, the complexity of AP-mMIMO (ZF) and AP-mMIMO (MP) are only $17.85 \%$ and $11.94 \%$ of that of MU-PNL-GDm, respectively. What is more, the proposed AP-mMIMO (MP) requires lower complexity than the classical ZF+DPD method, about $53\%$ of the complexity required when $\tau=5$. Another benefit of our approach is related to the fact that it is a data-driven one. So, its performance do not depend on the availability of an accurate PA model. However, the performance of ZF+DPD and MU-PNL-GDm methods deteriorates when there is a model deficiency in capturing the actual PA properties.   

In addition, we note that when $\tau$ is small ($\leq 10$), the AP-mMIMO (MP) scheme requires lower complexity than the AP-mMIMO (ZF) one. However, this latter  performs better, in terms of complexity, when $\tau$ becomes larger.  

\section{Conclusion}
\label{sec:conclu}
In this paper, we introduced a novel DL Autoprecoder based mMIMO downlink transmission, which is capable to learn a precoder and a decoder to eliminate the MUI and compensate the PA impairments over varying fading channels. Numerical results clearly showed the ability of the proposed AP-mMIMO to provide competitive performance with a much lower computational complexity compared to existing literature. Interesting findings of this paper include: 1) The adopted two-stage precoding scheme  makes the proposed AP-mMIMO suitable for varying fading channels. So, only the linear precoding is adapted once the channel changes.  2) The NN-precoder/decoder can be trained off-line and do not depend on the instantaneous channel estimates. 3) The AP-mMIMO (MP) is recommended for fast varying channel scenarios while the AP-mMIMO (ZF) performs better for slow varying channels.

To our knowledge, this work represents the first exploration of 
the potential of
DL in dealing with PA non-linearities in massive MU-MIMO downlink systems over varying fading channels, offering interesting insights for the design of energy-efficient massive MU-MIMO based 6G systems.
As a future work, the ideas presented here could be extended to study the capability of end-to-end DL approaches for intelligent reflecting surface (IRS), a promising solution for 6G communications.

\section*{Acknowledgment}
A part of this work was done when Rafik Zayani was working with Innov'COM lab at Sup'Com/Carthage University.

\bibliographystyle{unsrt}
\bibliography{reference}

\end{document}